\documentclass[twocolumn,prl,superscriptaddress,floatfix]{revtex4}
\usepackage{graphicx}
\usepackage{xcolor}
\usepackage{epsfig}
\usepackage{rotating}
\usepackage{amsmath}
\usepackage{amsfonts}
\usepackage{amssymb}
\usepackage{enumerate}
\usepackage{longtable}
\usepackage{appendix}
\usepackage{dcolumn}
\usepackage{bm}
\DeclareMathOperator{\Tr}{Tr}
\usepackage{hyperref}
\usepackage{ulem}

\begin{document}

\title{Thermodynamic behavior of modified integer-spin Kitaev models on 
the honeycomb lattice}

\author{Owen Bradley}
\affiliation{Department of Physics, University of California Davis, 
California 95616, USA}

\author{Jaan Oitmaa}
\affiliation{School of Physics, The University of New South Wales,
Sydney 2052, Australia}

\author{Diptiman Sen}
\affiliation{Center for High Energy Physics and Department of Physics, 
Indian Institute of Science, Bengaluru 560012, India}

\author{Rajiv R. P. Singh}
\affiliation{Department of Physics, University of California Davis, 
California 95616, USA}

\date{\rm\today}

\begin{abstract}
We study the thermodynamic behavior of modified spin-$S$ Kitaev models 
introduced by Baskaran, Sen, and Shankar [Phys.~Rev.~B {\bf 78}, 115116 
(2008)]. These models have the property that for half-odd-integer spins their 
eigenstates map on to those of spin-1/2 Kitaev models, with well-known 
highly entangled quantum spin-liquid states and Majorana fermions. For 
integer spins, the Hamiltonian is made out of commuting local operators. 
Thus, the eigenstates can be chosen to be completely unentangled between 
different sites, though with a significant degeneracy for each eigenstate.
For half-odd-integer spins, the thermodynamic properties can be related to the 
spin-1/2 Kitaev models apart from an additional degeneracy. Hence we focus 
here on the case of integer spins. We use transfer matrix methods, high-temperature expansions, and Monte Carlo simulations to study the thermodynamic 
properties of ferromagnetic and antiferromagnetic models with spin $S=1$ and 
$S=2$. Apart from large residual entropies, which all the models have, we
find that they can have a variety of different behaviors. Transfer matrix 
calculations show that for the different models, the correlation lengths can 
be finite as $T\to 0$, become critical as $T\to 0$ or diverge exponentially 
as $T\to 0$. The $Z_2$ flux variable associated with each 
hexagonal plaquette saturates at the value $+1$ as $T\rightarrow0$ in 
all models except the $S=1$ antiferromagnet where the mean flux remains zero 
as $T\to 0$. We provide qualitative explanations for these results.
\end{abstract}


\maketitle

\section{I. Introduction}

In the study of quantum spin systems~\cite{Balents_nature}, Kitaev materials have emerged as a very active field of 
research~\cite{balents2,trebst,kitaevm1,kitaevm2,banerjee}. Following Kitaev's 
introduction of 
the honeycomb lattice model with bond-dependent Ising interactions for 
different spin components~\cite{kitaev} and subsequent theoretical work~\cite{feng,nussinov1,nussinov2,baskaran1,
vidal,lee,willans,santosh,price,sela}, Jackeli and Khaliullin showed how 
such exchanges can be realized in real materials~\cite{khaliullin}. Since 
then, a number of materials have been proposed where such interactions
are dominant~\cite{kitaevm1,kitaevm2,banerjee}. Possible observation of fractionally quantized thermal Hall 
effect~\cite{matsuda} has made these systems central
in the search for quantum spin-liquid phases of matter.

While much interest has justifiably focused on spin-1/2 Kitaev materials, it 
has been shown that Kitaev models with arbitrary spin retain many interesting 
properties~\cite{baskaran,chandra,samarakoon,koga1,rous,suzuki,minakawa,koga2,
stavropoulos,xu,khait}. 
For arbitrary spin, the system is a classical spin-liquid at intermediate 
temperatures, with only very short-range spin correlations and an extensive
classical degeneracy. There are conserved flux variables associated with each elementary hexagonal plaquette.
The lack of exact solubility of the models for spin greater than half means that the
ground state is not known. One very interesting issue is the possibility that the property of half-odd-integer and
integer spins can be very different at low temperatures. Like the famous Haldane chain problem in one dimension, 
half-odd-integer spin models may have gapless excitations while integer spins may be gapped. Increasingly, many studies have focused on material realization of higher-spin Kitaev models~\cite{stavropoulos,xu,hickey}.

In Ref.~\cite{baskaran}, Baskaran, Sen, and Shankar (BSS) introduced a simpler 
model which we call a modified Kitaev model. This model 
shares some key features with the Kitaev model. It is defined by replacing 
the spin operators $S_i^\alpha$ in the
Kitaev model by the operators $\tau_i^\alpha = e^{i\pi S_i^\alpha}$. 
The model has Ising couplings between different components of the $\tau$ 
variables on different bonds, just like in the Kitaev model. 
This model continues to have conserved local flux variables on each hexagonal 
plaquette of the honeycomb lattice. For each half-odd-integer spin, the model is 
equivalent to many copies of the spin-1/2 Kitaev model. Thus its eigenstates are highly entangled and support Majorana fermion excitations. In contrast, for integer-spins, the basic $\tau$ operators at a site commute with each other. Thus all eigenstates can be chosen to be a product state from site to site with no intersite entanglement. However, the system remains highly degenerate, which leaves open the possibility of constructing entangled eigenstates via superposition of the many degenerate states. 
The modified Kitaev model is an interesting statistical model in its own right and the goal of this paper is to understand its thermodynamic behavior.

The Hamiltonian for the modified Kitaev model is~\cite{baskaran}:
\begin{equation}
H = J \left(\sum_{\langle i,j \rangle} \tau^x_i \tau^x_j + \sum_{(i,k)}\tau^y_i \tau^y_k + \sum_{[i,l]}\tau^z_i \tau^z_l\right), \label{Hamiltonian}
\end{equation} 
where $\tau^\alpha_i = e^{i\pi S^\alpha_i}$ and $\langle i,j \rangle$, $(i,k)$, and $[i,l]$ denote nearest-neighbor pairs which have a bond pointing along 
the $x$, $y$, and $z$ bond directions of the honeycomb lattice, respectively. $S^\alpha_i$ ($\alpha = x,y,z$) are the spin operators (defined at each site $i$) which are $(2S+1) \times (2S+1)$ matrices. Since the spin operators $S^x$, $S^y$, and $S^z$ satisfy $e^{i \pi S^a} e^{i \pi S^b} = (-1)^{2S} e^{i \pi S^b} e^{i \pi S^a}$ (for $a \neq b$), for integer spins the $\tau^\alpha$ operators commute and hence can be simultaneously diagonalized. As shown in Appendix A, the resulting matrices have diagonal elements $\pm 1$, and the model becomes a classical statistical model. 

In this work we study the integer spin BSS models using high-temperature expansions,
transfer matrices, and Monte Carlo simulations. We find a very rich finite-temperature thermodynamic behavior, which can be contrasted with the known results for the Kitaev model. In the Kitaev model, the intermediate temperature physics is dominated by
an entropy plateau at exactly half the maximum entropy regardless of the spin and ferromagnetic vs antiferromagnetic coupling
in the model. The existence of such plateaus is an important indicator of a low-energy subspace from which the quantum spin-liquid states emerge. In Appendix B, we provide a semiclassical
explanation for this residual entropy of large-spin Kitaev models. 
In contrast to the Kitaev model, for the BSS model ferromagnetic and antiferromagnetic
models behave very differently, and between the $S=1$ and $S=2$ ferromagnetic and antiferromagnetic cases, we find several classes of behaviors. For the $S=1$ antiferromagnet, the system stays disordered as $T\to 0$ with zero net flux and a short correlation length. The $S=2$ antiferromagnet appears to have an exponentially diverging correlation length as $T\to 0$ on finite-width cylinders. It is known that in the one-dimensional transverse field Ising model, as the magnetic field is varied (near $T=0$) there is analogous behavior of the correlation length in the low and high field regimes. There is a disordered gapped phase at $T=0$ with short-range correlations at high fields, and an ordered gapped phase in which the correlation length becomes exponentially large as $T\to 0$. In this sense, the antiferromagnetic models for $S=1$ and $S=2$ appear to be
gapped and on two sides of a $T=0$ order-disorder transition. 
In contrast the $S=1$ and $S=2$ ferromagnetic models appear critical with correlation length scaling with the width
of the cylinder, presumably implying critical correlations and gapless excitations in the thermodynamic limit. This shows that the modified Kitaev models have a wide variety of thermodynamic behavior at intermediate temperatures. 

In Sec.~II we discuss the $\tau$ operators for $S=1$ and $S=2$ and some general properties of our model. In Sec.~III we discuss the transfer matrix method 
for a finite-width cylindrical strip, our Monte Carlo simulations, and the method of high-temperature series expansions. In Sec.~IV the numerical results for the model are presented and we discuss some analytical insights into these results. We consider the integer-spin models only in this work. Finally in Sec.~V we present our conclusions, open questions and the relevance of these studies to the Kitaev family of materials. 

\section{II. Modified Kitaev Model}

\begin{figure}[htb!]
\centering
\includegraphics[width=\columnwidth]{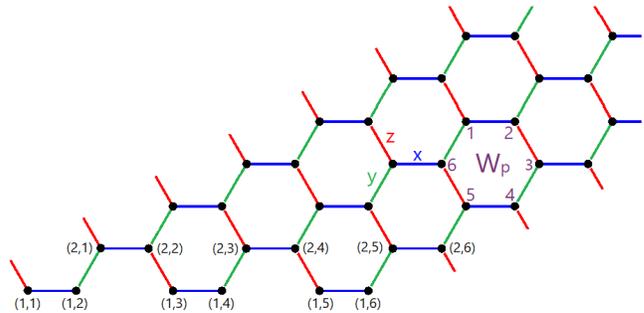}
\caption{Geometry of the honeycomb lattice (shown for a semi-infinite lattice 
$n=6$ sites in width), with the $x$, $y$, and $z$ bond directions indicated. 
Each site may be labeled $(\alpha, \beta)$ where $\alpha$ denotes the row 
number and $\beta = 1, \ldots, n$ denotes the position of the site along its 
row.} \label{lattice} \end{figure}

For each hexagon in the lattice (with sites labeled $1, \ldots, 6$ as shown in Fig.~\ref{lattice}) one can define the plaquette flux operator 
\begin{equation}
W_p = e^{i \pi (S_1^z + S_2^y + S_3^x + S_4^z + S_5^y + S_6^x)} = \Pi_{i=1}^6 \tau_i^{\alpha_{out}}, \label{FluxDef}
\end{equation}
where $\alpha_{out}$ is the direction pointing out of the loop at site $i$. As shown in Ref.~\cite{baskaran}, the $W_p$ operators both commute with the Hamiltonian and each other and have eigenvalues equal to $\pm 1$. Hence the model has conserved $Z_2$ flux variables on each hexagonal plaquette of the honeycomb lattice. For integer spin, this conservation is trivial as each $\tau_i$ commute with each other. Nevertheless, the thermal expectation value of these flux variables is strongly correlated with the thermodynamic behavior of the model, as we will show here, resembling that in the original Kitaev models.
This suggests that the modified Kitaev models preserve some aspects of the Kitaev model thermodynamics despite their simplicity.

The $\tau^\alpha$ ($\alpha = x, y, z$) operators can be simultaneously diagonalized, giving three diagonal matrices of dimension $(2S+1)\times(2S+1)$. As shown in Appendix A, for $S=1$ this gives three possibilities for $(\tau^x, \tau^y, \tau^z)$ at each site: $(1, -1, -1)$, $(-1, 1, -1)$, and $(-1, -1, 1)$. 
For $S=2$, the diagonalized tau operators yield five possibilities for $(\tau^x, \tau^y, \tau^z)$ per site, which have $(\tau^x, \tau^y, \tau^z) = (-1, -1, 1)$, $(-1, 1, -1)$, $(1, 1, 1)$, $(1, -1, -1)$, and $(1, 1, 1)$. Note that the five triplets contain those found for $S=1$, plus an additional pair for which $\tau^x = \tau^y = \tau^z = 1$.

\subsection{A: Zero-temperature entropy for the spin-1 ferromagnetic model}

We will derive here an exact result for the ground-state entropy per site for the spin-1 
ferromagnetic model where the nearest-neighbor interactions are taken to be 
of the form $\tau_m^a \tau_n^a$, where $\tau_m^a = e^{i \pi S_m^a}$ at site
$m$. For the spin-1 case, we find that
\begin{equation} \tau_m^x ~\tau_m^y ~\tau_m^z ~=~ 1, \label{cond1} 
\end{equation}
for each $m$, and the three possible states at each site have $(\tau^x,\tau^y,
\tau^z) = (1,-1,-1)$, $(-1,1,-1)$, and $(-1,-1,1)$ in a suitably defined basis.

The Hamiltonian of the model is given by
\begin{equation} H ~=~ J ~\sum_{mn} \tau_m^a ~\tau_n^a, \label{ham3} 
\end{equation}
where the sum goes over nearest-neighbor pairs of sites $(mn)$ of the 
honeycomb lattice, and $a$ depends on the direction of the bond joining 
$(mn)$. The ferromagnetic case corresponds to $J < 0$. Now, 
each state at site $m$ has one of the $\tau_m^a$'s equal to $+1$; hence
the ground state for the bond involving $\tau_m^a = +1$ must have
its neighboring $\tau_n^a$ also equal to $+1$. Let us call this bond a 
dimer. All the nondimer bonds have both $\tau_m^a$ and $\tau_n^a$ equal 
to $-1$, which is also the ferromagnetic ground state of those bonds.
The total number of ground states of the ferromagnetic model is therefore
given by the number of dimer coverings of the honeycomb lattice. In the 
thermodynamic limit in which the number of sites $N \to \infty$, it is 
known~\cite{kasteleyn,baxter,wu} that the number of dimer coverings is 
given by $(1.381)^{N/2}$. Hence the ground-state entropy per site is given by 
$(1/2) \ln (1.381) \simeq 0.161$.

\subsection{B: Ground state energy per site equal to $-3/2$ implies that the
flux per hexagon is $+1$}

The identity in Eq.~\eqref{cond1} holds for both spin-1 and spin-2 models.
For the cases of spin-1 ferromagnet, spin-2 ferromagnet, and spin-2 
antiferromagnet, it is found that the ground-state energy per site is
equal to $-3/2$ (for $|J| = 1$). This means that every bond simultaneously has minimum energy.
We will now show that in all these cases, 
the flux in each hexagon must be equal to $+1$. The flux is defined as 
\begin{equation} W_p ~=~ \tau_1^z ~\tau_2^y ~\tau_3^x~ \tau_4^z ~\tau_5^y ~
\tau_6^x \label{wp1} \end{equation}
(see Fig.~\ref{lattice}). Equation \eqref{cond1} implies that
$\tau_1^z = \tau_1^x \tau_1^y$, and so on. Hence Eq.~\eqref{wp1} can be
re-written as
\begin{equation} W_p ~=~ (\tau_1^x \tau_2^x)~ (\tau_2^z \tau_3^z)~ (\tau_3^y 
\tau_4^y)~ (\tau_4^x \tau_5^x)~ (\tau_5^z \tau_6^z)~ (\tau_6^y \tau_1^y)~. 
\label{wp2} \end{equation}
Each of the terms in parentheses in Eq.~\eqref{wp2} corresponds to one of the 
interaction terms in the Hamiltonian, and each such term is equal to $+1$ or 
$-1$ in the ground state depending on whether the model is ferromagnetic or 
antiferromagnetic, provided that the ground-state energy per spin attains 
the value $-3/2$. We therefore see that $W_p$ must be equal to $+1$ in 
every hexagon in such ground states.

\section{III. Methods}

\subsection{A. Transfer Matrix Method}

One approach we employ to study the thermodynamic properties of our model is the transfer matrix method, which we now describe in detail. For spin-1, each site has three possible spin states $\sigma=\{1,2,3\}$, and we have $(\tau^x_i, \tau^y_i, \tau^z_i) = (1,-1,-1)$, $(-1,-1,1)$ or $(-1,-1,1)$ if site $i$ is in state 1, 2, or 3 respectively. Consider a honeycomb lattice formed by stacking $N$ rows, each consisting of $n$ sites. There are $3^n$ possible configurations on each row. The total energy of the lattice obtained by Eq.~\eqref{Hamiltonian} is denoted $E\{\sigma_i\}$, which depends on the state at each site. The partition function is given by
\begin{equation}
Z = \sum_{\{ \sigma_i \}} e^{-\beta E\{\sigma_i\}}, \label{partition}
\end{equation}
where the sum is over all possible spin states at each site. Labeling the 
rows as $1, 2, \ldots, N$, this can be written as
\begin{equation}
Z = \sum_{\{ \sigma_1 \}} \sum_{\{ \sigma_2 \}} \ldots \sum_{\{ \sigma_N \}} e^{-\beta E(\{\sigma_1\}, \{\sigma_2\},\ldots,\{\sigma_N\})}, \label{partition2}
\end{equation}
where $\sum_{\{ \sigma_1 \}}$ denotes a sum over all $3^n$ configurations of row 1, etc. The transfer matrix method relies on the fact that one can construct a particular matrix $T$ such that the partition function can be written as
\begin{equation}
Z = \sum_{\{ \sigma_1 \}} \sum_{\{ \sigma_2 \}} \ldots \sum_{\{ \sigma_N \}} T_{1;2} T_{2;3} T_{3;4} \ldots T_{N-1;N} T_{N;1}, \label{partition3}
\end{equation}
where $T_{A;B}$ is a $3^n \times 3^n$ matrix containing contributions to the total energy arising from a \textit{pair} of rows (labeled A and B), with row B directly above row A, and we have assumed periodic boundary 
conditions by writing $T_{N;1}$. The partition function now becomes
\begin{equation}
Z = \sum_{\{ \sigma_1 \}} T^N_{1;1} = \Tr(T^N), \label{partition4}
\end{equation}
where $T$ is a $3^n \times 3^n$ matrix. The elements of T are given by
\begin{equation}
T(i,j) = \exp[-\beta(E_1 + E_2 + E_3 + E_4)], \label{elements}
\end{equation}
where the energy terms are as given in Eqs.~(\ref{E1term})--(\ref{E4term})
below. To find the matrix element $T(i,j)$ we consider a pair of rows labeled 1 and 2, with row 2 directly above row 1. Let rows 1 and 2 have configurations $i$ and $j$, respectively, where $i$ and $j$ denote one of $3^n$ possible row configurations. For clarity, let us label each lattice site by $(\alpha, \beta)$, where $\alpha$ denotes the row number of the site, and $\beta=1,2,\ldots,n$ denotes the position of the site along its row as shown in Fig.~\ref{lattice}. The energy terms in Eq.~\eqref{elements} are then given by 
\begin{align}
E_1 &= \frac{1}{2} \left( \tau^x_{1,1} \tau^x_{1,2} + \tau^x_{1,3} \tau^x_{1,4} + \ldots + \tau^x_{1,n-1} \tau^x_{1,n} \right), \label{E1term} \\
E_2 &= \frac{1}{2} \left( \tau^x_{2,1} \tau^x_{2,2} + \tau^x_{2,3} \tau^x_{2,4} + \ldots + \tau^x_{2,n-1} \tau^x_{2,n} \right), \label{E2term} \\
E_3 &= \tau^y_{1,2} \tau^y_{2,1} + \tau^y_{1,4} \tau^y_{2,3} + \tau^y_{1,6} \tau^y_{2,5} + \ldots + \tau^y_{1,n} \tau^y_{2,n-1}, \label{E3term} \\ 
E_4 &= \left( \tau^z_{1,3} \tau^z_{2,2} + \tau^z_{1,5} \tau^z_{2,4} + \ldots + \tau^z_{1,n-1} \tau^z_{2,n-2} \right) + \tau^z_{1,1} \tau^z_{2,n}, \label{E4term}
\end{align}
The factors of $1/2$ in $E_1$ and $E_2$ are included to avoid double counting interactions from the $x$ bonds. The last term in $E_4$ is included due to the periodic boundary condition.

Since the trace of $T^N$ is independent of the basis used, if we consider the basis where $T$ is diagonal, then $T^N =$ diag$(\lambda^N_1, \lambda^N_2, \lambda^N_3, \ldots, \lambda^N_{3^n})$, where $\lambda_i$ are the eigenvalues of $T$. Letting $\lambda_1$ denote the largest eigenvalue, the partition function can now be written as
\begin{align}
Z &= \lambda^N_1 + \lambda^N_2 + \lambda^N_3 + \ldots \\
&= \lambda^N_1 \left[1 + \left(\frac{\lambda_2}{\lambda_1}\right)^N + \left(\frac{\lambda_3}{\lambda_1}\right)^N + \ldots \right].
\end{align}
In the limit $N \rightarrow \infty$ (i.e., a semi-infinite lattice) we therefore have $Z = \lambda^N_1$. The free energy can then be found from $ F = -k_B T \ln Z = -k_B T N \ln(\lambda_1)$, and since the total number of sites is $N_{tot} = N \times n$, the free energy per site is given by
\begin{equation}
f = \frac{F}{N_{tot}} = - k_B T ~\frac{\ln(\lambda_1)}{n}. \end{equation}
Thermodynamic quantities such as entropy, specific heat, and internal energy can then be obtained by taking suitable derivatives of the free energy. The correlation length can also be found from the ratio of the two largest eigenvalues of the transfer matrix
\begin{align}
\xi &= \frac{1}{\ln(\lambda_1 / \lambda_2)}. \label{xiequation}
\end{align}

\begin{figure*}[htb!]
\centering
\includegraphics[width=1.7\columnwidth]{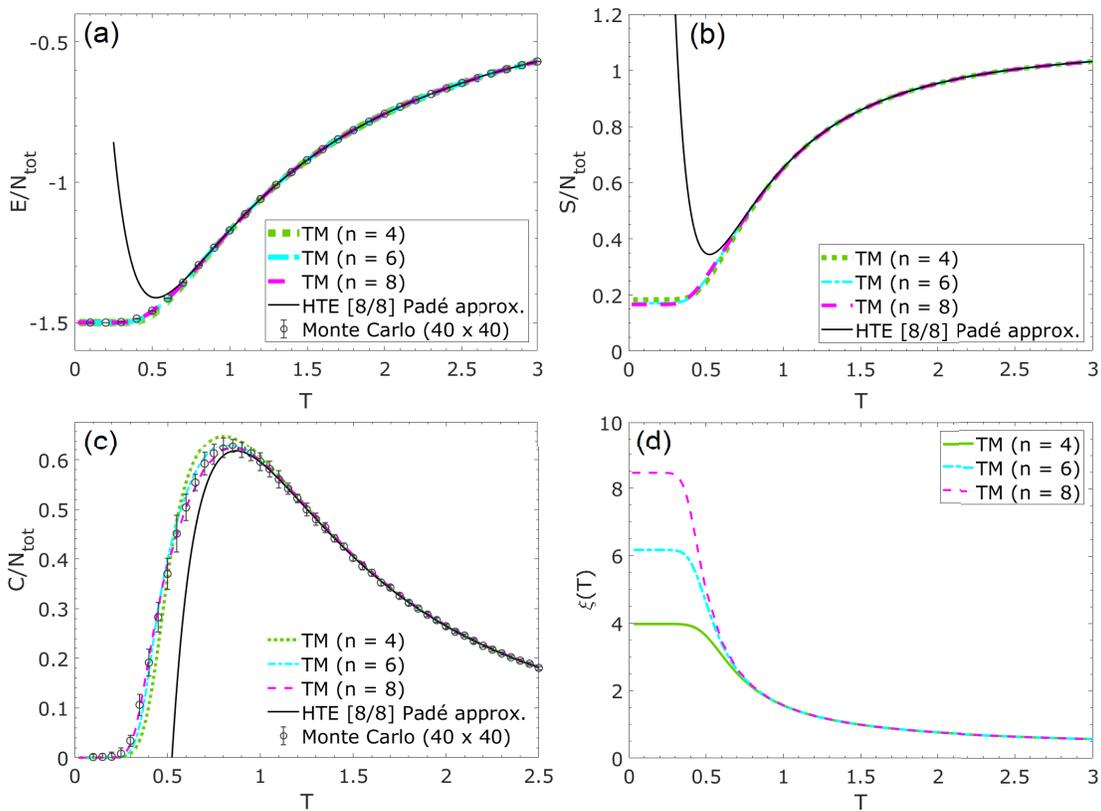}
\caption{$S=1$ ferromagnetic model $(J<0)$: Transfer matrix (TM), 
high-temperature expansion (HTE) and Monte Carlo results are shown for 
(a) energy per site, (b) entropy per site, (c) specific heat per site, and 
(d) correlation length, as a function of temperature. TM results are shown 
for systems $n=4$, $6$, and $8$ sites wide, and HTE data are extrapolated using 
a Pad\'{e} approximant of order $[8,8]$. Monte Carlo data are shown for 
lattices of size $40\times40$.} \label{S1FM_Plot} \end{figure*}

\subsection{B. Monte Carlo Method}

We also employ a Monte Carlo procedure to study the thermodynamic properties of the model. This allows us to verify our transfer matrix and high-temperature
series expansion results and also investigate the thermodynamic behavior of the flux variable $W_p$ defined previously. For both the spin-1 and spin-2 cases, the $\tau^\alpha$ operators commute and can be simultaneously diagonalized with diagonal entries equal to $\pm 1$. Thus for integer spin, our model is akin to an Ising spin model which can be studied using classical Monte Carlo methods.

For spin-1, there are three possible states per site, whereas for spin-2 there are five possible states. We use the Metropolis-Hastings algorithm to perform local updates of these states. That is, for each Monte Carlo sweep, a change in the state at each site is proposed in sequence, and one accepts the move if the energy difference $\Delta E < 0$, otherwise the move is accepted with probability $e^{-\beta \Delta E}$. Simulations were performed for honeycomb lattices with periodic boundary conditions for systems with up to $40 \times 40$ sites. We present Monte Carlo results for both the mean energy and specific heat per site, and also the mean flux per plaquette as a function of temperature, which is evaluated by averaging the mean flux over sampled configurations, i.e.,
\begin{equation} \langle W_p \rangle = \frac{1}{N_{meas}}\frac{1}{N_p} 
\sum_{i=1}^{N_{meas}}\sum_{p=1}^{N_p} W_p, \label{meanflux} \end{equation}
where $N_{meas}$ is the number of measurement sweeps performed (at least $10,000$ in our simulations), and $N_p=N/2$ is the number of hexagonal plaquettes in the lattice, which is equal to half the total number of sites.

We also compute a second moment correlation length $\xi_2$ corresponding to state-state correlations, defined by
\begin{equation}
\xi_2^2 = \frac{1}{2d} \frac{\sum_{ij} r_{ij}^2 (\langle \sigma_i \sigma_j \rangle - \langle \sigma_i \rangle \langle \sigma_j \rangle )}{\sum_{ij} (\langle \sigma_i \sigma_j\rangle - \langle \sigma_i\rangle \langle \sigma_j\rangle)},
\end{equation}
where $\sigma_i$ denotes the state at site $i$, $r_{ij}$ is the distance between two sites in the lattice, and $d=2$ is the number of dimensions. We map the states onto vectors in two dimensions, i.e., for spin-1 the three states are mapped to $(1,0)$, $(-1/2,\sqrt{3}/2)$, and $(-1/2,-\sqrt{3}/2)$. The correlation $\sigma_i \sigma_j$ is then defined as the scalar product of the vectors corresponding to the states at sites $i$ and $j$, and our assignment is symmetric between the three possible states.

\subsection{C. High Temperature Series Expansion Method}

The partition function for any interacting many-particle system is given by 
\begin{equation}
Z=\Tr\{e^{-\beta H}\}. 
\end{equation}
The quantity
\begin{equation}
A \equiv (1/N) \ln Z
\end{equation}
can be obtained via a linked-cluster method as 
\begin{equation}
A_{bulk} = \Sigma_g c_g \tilde{A}_g
\end{equation}
where the sum is over a set $\{g\}$ of finite clusters of increasing size, $c_g$ is an embedding constant, and $\tilde{A}_g$ is the ``reduced" value of $\ln Z$ for cluster $g$, with all of the subcluster contributions subtracted off. Technical details are explained in Ref.~\cite{oitmaa_book}. In the present classical case, where all operators commute, we can use the simpler form, 
\begin{equation}
Z = \Sigma_n e^{-\beta E_n} = \Sigma_r g_r e^{-\beta E_r}
\end{equation}
where the first sum is over all states $\{n\}$, which can be grouped into a sum over distinct eigenstates $\{r\}$ with $g_r$ being the degeneracy. For any finite cluster $g$ the partition function is then a finite, fairly small, sum of weighted exponentials, which can be expanded to give a series in powers of $\beta = 1/k_B T$, from which $\ln Z$ is easily obtained. The bulk series will then be correct to an order determined by the number of bonds in the largest cluster. For $S = 1$ we have computed a series to order $\beta^{16}$, using a set of 17060 clusters with up to 16 bonds. For $S = 2$ a series to order $\beta^{14}$ was obtained, requiring 3453 distinct clusters. We do not list the series coefficients here, but they are available in Supplemental Material \cite{supplemental}. From the series for $(1/N) \ln Z$ we obtain series for entropy, energy and specific heat using standard thermodynamic relations. The series are then evaluated using standard Pad\'{e} approximant methods.


\begin{figure*}[t!]
\centering
\includegraphics[width=1.7\columnwidth]{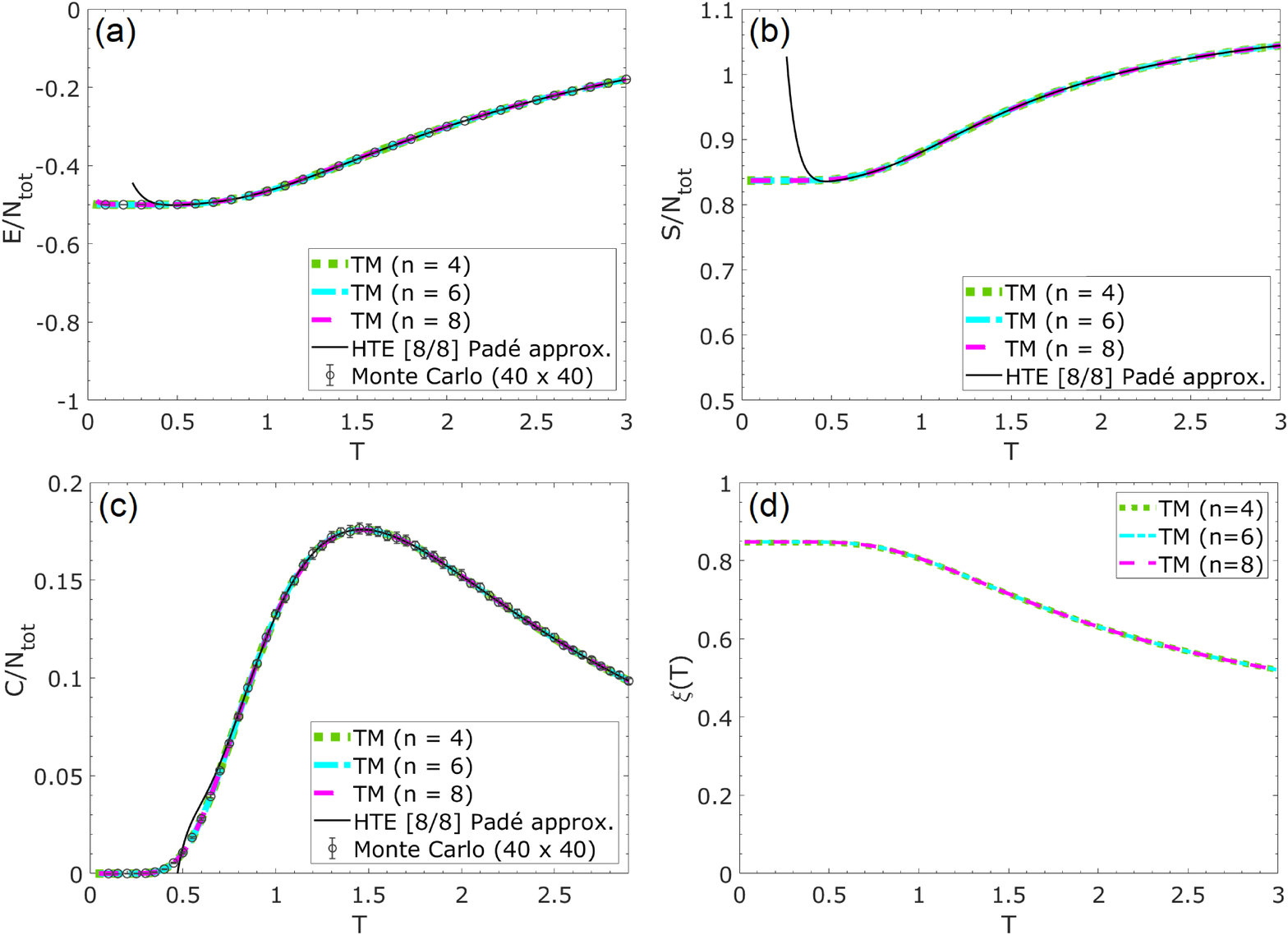}
\caption{The $S=1$ antiferromagnetic model $(J>0)$: TM, HTE, and Monte Carlo results are shown for 
(a) energy per site, (b) entropy per site, (c) specific heat per site, and 
(d) correlation length, as a function of temperature. TM results are shown 
for systems $n=4$, $6$ and $8$ sites wide, and HTE data are extrapolated using 
a Pad\'{e} approximant of order $[8,8]$. Monte Carlo data are shown for 
lattices of size $40\times40$.} \label{S1AF_Plot} \end{figure*}

\section{IV. Results and Discussion}

\subsection{A: Thermodynamics of Spin-1 model}

In this work we have studied the thermodynamic properties of the modified Kitaev model for $S=1$ and $S=2$, for both ferromagnetic ($J<0$) and antiferromagnetic ($J>0$) couplings, using high-temperature series expansions, transfer matrices, and Monte Carlo simulations (note that we take $|J|=1$ throughout). We first present results for the entropy, specific heat, energy and correlation length for the ferromagnetic $S=1$ case. Since the transfer matrix procedure involves finding the eigenvalues of a $3^n \times 3^n$ matrix we are computationally limited to studying lattices infinite in one direction but no more than $n=8$ sites wide. Figure \ref{S1FM_Plot}(b) shows the entropy per site calculated using the transfer matrix method for systems of width $n=4$, $6$ and $8$. There is a nonzero residual entropy as the temperature approaches zero, resulting from the large degeneracy of the ground state. With increasing lattice size, the residual entropy per site approaches $(1/2)\ln(1.381) \approx 0.161$ as $T\rightarrow0$, which is directly related to the number of dimer coverings of the honeycomb lattice as discussed in Sec.~II.~A. We also observe a plateau in $S(T)$ at the value of the residual entropy for $T \lesssim 0.3$ and find that the entropy per site approaches $\ln(3)$ in the high-temperature limit, as expected. We also plot the HTE result for the entropy for comparison. There is a close agreement between the transfer matrix and HTE data at high temperature, indicating that our result is accurate in the thermodynamic limit. However the HTE convergence breaks down before the onset of the low-temperature entropy plateau. 

The specific heat shows a single broad peak, with the HTE result in close agreement with the $n=8$ transfer matrix result in this region as shown in Fig.~\ref{S1FM_Plot}(c). In addition, our transfer matrix result for $C(T)$ agrees well with Monte Carlo data obtained for $40 \times 40$ lattices. Figure \ref{S1FM_Plot}(a) shows the ground-state energy per site is equal to $-3/2$, which corresponds to all bonds of the honeycomb lattice becoming satisfied as $T \rightarrow 0$. We also use Eq.~\eqref{xiequation} to find the correlation length $\xi$ from the ratio of the two largest eigenvalues of the transfer matrix. For the ferromagnetic $S=1$ case, $\xi$ scales with the lattice width as $T\rightarrow0$ indicating critical behavior, as shown in Fig.~\ref{S1FM_Plot}(d). 

For the antiferromagnetic $S=1$ case, there are several notable differences in the thermodynamic behavior compared to the ferromagnetic model, as shown in Fig.~\ref{S1AF_Plot}. Firstly, there is a much larger residual entropy of approximately 0.837, with a plateau present for $T \lesssim 0.5$. Since there is no configuration for which all antiferromagnetic couplings in the lattice are satisfied, the ground-state energy per site is also greater and attains the value $-1/2$. Thus the system remains disordered as $T\rightarrow 0$. This energy per site corresponds to having $2/3$ of the bonds in the lattice satisfied and $1/3$ unsatisfied (e.g., those in a particular bond direction $x$, $y$, or $z$) in the ground state. In contrast to the ferromagnetic case, our transfer matrix results exhibit almost no dependence on the system width $n$ for any of the lattice sizes studied, and there is a closer agreement with HTE data for specific heat at temperatures below the peak. The correlation length also remains short as $T\rightarrow 0$ independent of $n$, and does not scale with system width. As in the ferromagnetic case, our transfer matrix results for $E(T)$ and $C(T)$ agree well with Monte Carlo simulations on $40 \times 40$ lattices, indicating that the finite-width cylinders studied yield quite accurate results for our purposes.

\begin{figure}[t!]
\centering
\includegraphics[width=\columnwidth]{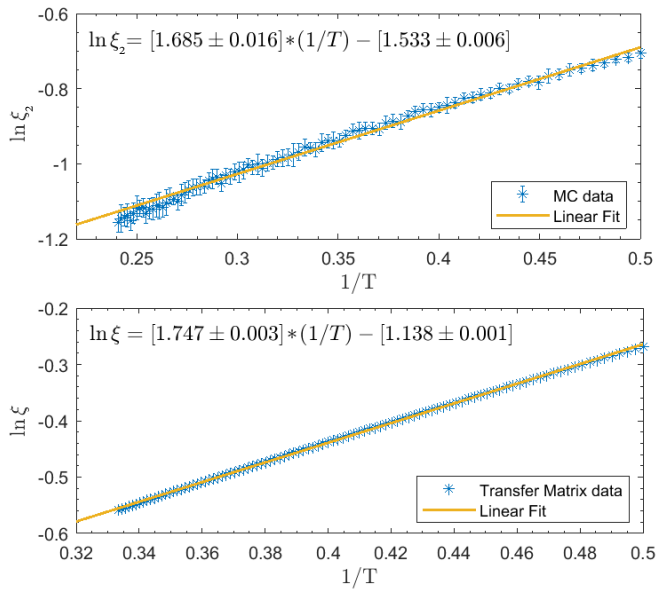}
\caption{Top: Monte Carlo result for the state-state correlation function for the $S=1$ ferromagnetic model ($6 \times 6$ lattice). $\ln \xi_2$ is plotted as a function of $1/T$ along with a linear least-squares fit with slope $1.685\pm0.016$. Bottom: Transfer matrix result for the correlation length for a system $n=6$ sites wide, again for the $S=1$ ferromagnet. $\ln \xi$ is plotted as a function of $1/T$ along with a linear least-squares fit with slope $1.747\pm0.003$.} \label{Correlation_Plot} \end{figure}

\textbf{Identification of the growing correlations:} The transfer matrix results do not tell us which correlation length is the largest in the system. In the Monte Carlo simulations, we have examined several correlation functions, including the spin-spin correlations involving the
variable $\tau_i$, the $Z_2$ flux-flux correlation functions for different hexagons and the
correlation between the occupation of the (2S+1) states of the system at different sites. Only the latter grows, 
as temperature is lowered, in a manner similar to the correlation length found from the transfer matrix calculations. As shown in Fig.~\ref{Correlation_Plot} for the $S=1$ ferromagnet, $\xi(T)$ from transfer matrix calculations behaves similarly to the state-state correlation length $\xi_2$ obtained in the Monte Carlo simulations. Plotting $\ln(\xi_2)$ as a function of $1/T$ for both yields approximately straight lines of comparable slope, suggesting they have similar functional forms. We note that, at low temperatures, we found it difficult to explore the behavior of the system with Monte Carlo simulations, even when incorporating some cluster moves which collectively change the states of sites belonging to one hexagon. Once the system enters the manifold of degenerate states, acceptance of the Monte Carlo moves goes rapidly to zero. Thus, at lower temperatures, we have to rely on the transfer matrix calculations.

\begin{figure*}[t!]
\centering
\includegraphics[width=1.7\columnwidth]{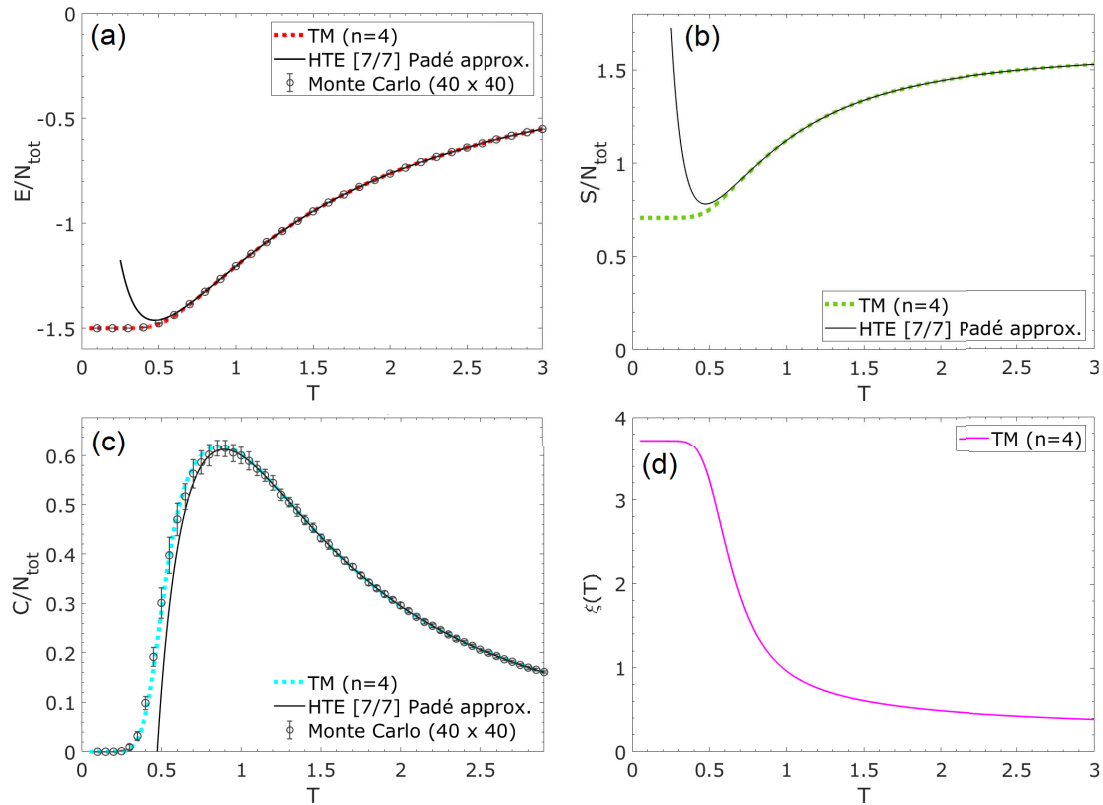}
\caption{The $S=2$ ferromagnetic model $(J<0)$: TM, HTE and Monte Carlo results are shown for 
(a) energy per site, (b) entropy per site, (c) specific heat per site, and 
(d) correlation length, as a function of temperature. TM results are shown 
for a system $n=4$ sites wide, and HTE data are extrapolated using a Pad\'{e} 
approximant of order $[7,7]$. Monte Carlo data are shown for lattices of size 
$40\times40$.} \label{S2FM_Plot} \end{figure*}

\subsection{B: Thermodynamics of Spin-2 model}

For both the ferromagnetic and antiferromagnetic $S=2$ models there are five possible states per site. The transfer matrix procedure thus limits us to studying finite-width cylinders no more than $n=4$ sites wide, since the method now requires finding the eigenvalues of a $5^n \times 5^n$ matrix. Nevertheless, we find the transfer matrix results closely agree with Monte Carlo data for $40\times40$ clusters at low temperature, and also with high-temperature series expansion results (which are formally defined in the thermodynamic limit) down to temperatures at which the specific heat peaks. 

\begin{figure*}[t!]
\centering
\includegraphics[width=1.7\columnwidth]{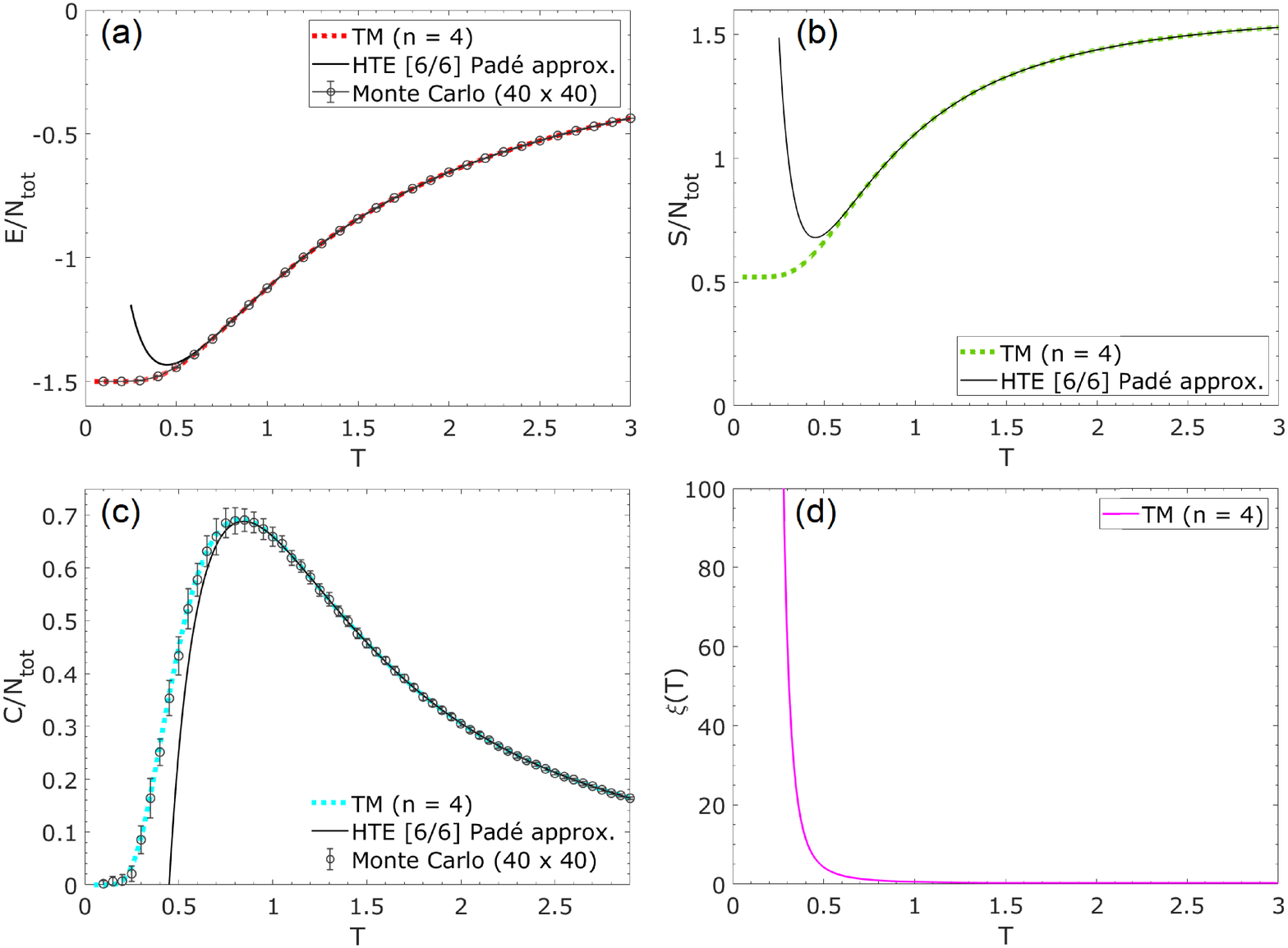}
\caption{The $S=2$ antiferromagnetic model $(J>0)$: TM, 
HTE, and Monte Carlo results are shown for 
(a) energy per site, (b) entropy per site, (c) specific heat per site and 
(d) correlation length, as a function of temperature. TM results are shown 
for a system $n=4$ sites wide and HTE data are extrapolated using a Pad\'{e} 
approximant of order $[6,6]$. Monte Carlo data are shown for lattices of size 
$40\times40$.} \label{S2AF_Plot} \end{figure*}

In Fig.~\ref{S2FM_Plot}, we show results for the entropy, specific heat, energy, and correlation length for the ferromagnetic $S=2$ case. As in all the models, there is a large ground-state degeneracy and a corresponding residual entropy per site as $T\rightarrow0$, which we find to be approximately $0.707$. In this model, the majority of ground states are those for which all sites have $(\tau_x, \tau_y, \tau_z)=(1,1,1)$; however, one can also find additional ground states by changing the state of sites belonging to closed loops in the lattice. In Sec.~IV C we provide a more thorough analytical explanation of this residual entropy value. Both the ferromagnetic $S=1$ and $S=2$ models have a ground-state energy per site of $-3/2$ indicating that all bonds in the lattice may be satisfied. Similarly to the $S=1$ case, the $S=2$ ferromagnet also appears critical with the correlation length limited by the width of the system, presumably implying critical correlations in the thermodynamic limit. As shown in Fig.~\ref{S2AF_Plot}(b), the antiferromagnetic $S=2$ model has a lower residual entropy per site of approximately 0.520. However, in contrast to the $S=1$ case, it is possible to satisfy all the bonds in the lattice leading to ground-state energy of $-3/2$ per site. The $S=2$ antiferromagnet also appears to have an exponentially diverging correlation length as the temperature is lowered. This is illustrated in Fig.~\ref{S2AF_Plot}(d), which shows the transfer matrix result for a semi-infinite lattice $n=4$ sites wide.

\subsection{C: Ground state entropy for the spin-2 ferromagnetic model}

For the spin-2 case, we know that there are five states at each site in which 
$(\tau^x,\tau^y, \tau^z) = (1,-1,-1)$, $(-1,1,-1)$, $(-1,-1,1)$, $(1,1,1)$,
and $(1,1,1)$ in a suitably defined basis. Note that there are two states both 
of which have $(\tau^x,\tau^y, \tau^z) = (1,1,1)$. For the ferromagnetic model,
a state in which
all sites have $(1,1,1)$ is clearly a ground state. Since this can happen in
two possible ways at each site, we see that the number of ground states 
be at least as large as $2^N$, for an $N$-site system. Hence the ferromagnetic model will have a ground state entropy per site which is at least as large as $\ln 2 \simeq 
0.693$. We will now argue that the entropy per site is a little larger than
this value.

Starting with one of the 
ground states in which all sites have $(1,1,1)$, suppose that we replace the 
states for the six sites in a particular hexagon by $(1,-1,-1)$, in which the
values $-1$ are along the bonds going around the hexagon while the value $1$
is along the bond which is directed away from the hexagon. This gives another
ground state since a bond which has $-1$ at both ends is satisfied for
a ferromagnetic interaction. The number of such states is, however, less
than $2^N$ by a factor of $2^6$, since we now have only one possible state
(instead of two) at each of the sites in that hexagon. Next, let the number 
of such hexagons where this replacement is made be $n$, where $n$ can go from 
zero to a maximum of $N/2$ (which is the number of hexagons for an $N$-site 
system). Actually, $n$ should be much less than $N/2$ since we cannot make 
such a replacements in two neighboring hexagons; we will see below that indeed 
$n \ll N/2$, so that two such hexagons are unlikely to be neighbors of each
other. The number of ways $n$ hexagons can be chosen out of $N/2$ is 
given by ${}^{N/2} C_n$. Hence the ground-state partition function is given by
\begin{equation} Z ~=~ 2^N ~\sum_{n=0,1,\cdots} ~\frac{1}{2^{6n}} ~{}^{N/2} 
C_n. \label{z1} \end{equation}

Introducing the variable $p=n/(N/2)$, we can rewrite Eq.~\eqref{z1} as
\begin{eqnarray} Z &=& 2^N ~\int_0^1 dp ~\exp [- ~\frac{pN}{2} \ln (2^6)
~+~ \frac{N}{2} ~\ln \frac{N}{2} \nonumber \\
&& ~-~ \frac{pN}{2} \ln \frac{pN}{2} ~-~ \frac{(1-p) N}{2} \ln 
\frac{(1-p) N}{2}], \label{z2} \end{eqnarray}
where we have used Stirling's formula for the factorial functions. We now 
find the maximum of the terms in the exponential in Eq.~\eqref{z2} as a 
function of $p$. This gives the condition 
\begin{equation} \frac{p}{1-p} ~=~ \frac{1}{2^6}. \end{equation}
Since $1/2^6 \ll 1$, this gives $p = 1/2^6$ to a good approximation.
Substituting this back in Eq.~\eqref{z2}, we find that
\begin{equation} Z ~=~ \exp [ N (\ln 2 ~+~ 1/2^7)]. \end{equation}
Hence the entropy per site is $\ln 2 + 1/2^7 \simeq 0.701$, which lies
below the value of $0.707$ found numerically.

The above argument was based on taking a closed loop and changing the
states for each site of that loop from $(1,1,1)$ to $(1,-1,-1)$. The smallest
possible closed loop forms a hexagon and this is what was assumed above. The 
next larger closed loop involves ten sites covering two neighboring hexagons. 
A similar argument as above will then give an additional contribution of
$1/2^{11} \simeq 0.0005$ to the entropy per site. Larger loops will give
additional contributions; however, these contributions will rapidly 
approach zero since a closed loop with $m$ sites will contribute $1/2^{m+1}$.
We have therefore only discussed here the contribution from the smallest 
closed loop which has $m=6$.

\subsection{D: Measurement of $Z_2$ Flux through a hexagonal plaquette}

\begin{figure}[t!]
\centering
\includegraphics[width=\columnwidth]{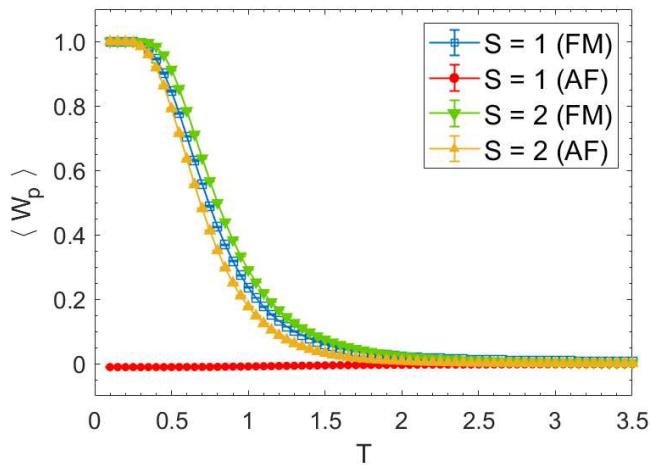}
\caption{The mean flux per hexagonal plaquette $\langle W_p \rangle$ as 
defined in Eq.~\eqref{meanflux} is plotted as a function of temperature for 
the $S=1$ ferromagnetic, $S=1$ antiferromagnetic, $S=2$ ferromagnetic, and 
$S=2$ antiferromagnetic models. The results are obtained from Monte Carlo 
simulations on lattices of size $40 \times 40$.} \label{Flux_Plot} \end{figure}

Our Monte Carlo simulations also allow us to study the thermodynamic behavior of the $Z_2$ flux variable defined in Eq.~\eqref{FluxDef}, which we perform on lattices of up to $40\times40$ sites with periodic boundary conditions. In Fig.~\ref{Flux_Plot}, we show the temperature dependence of the mean flux $\langle W_p \rangle$ through a hexagonal plaquette of the honeycomb lattice, for both the ferromagnetic and antiferromagnetic $S=1$ and $S=2$ models. In all cases, $\langle W_p \rangle \rightarrow 0$ as $T\rightarrow \infty$ as expected, since if the states at each site occur with equal likelihood, then fluxes of $+1$ and $-1$ occur with equal probability throughout the lattice. For the $S=1$ ferromagnetic case, and both the $S=2$ ferromagnetic and antiferromagnetic cases, we have seen that the ground-state energy per site is $-3/2$. As discussed in Sec.~II B, this implies that the flux per hexagon must become $+1$. For these cases, we indeed find that the mean flux saturates at this value for around $T \lesssim 0.5$, which corresponds to the temperature at which the ground-state energy is attained. Moreover, in these cases, the temperature at which $\langle W_p \rangle$ grows rapidly ($T \approx 1$) coincides with the peak in specific heat, which resembles the original spin-1 Kitaev model with ferromagnetic interactions \cite{koga1}. In contrast, the $S=1$ antiferromagnetic model, which has a ground-state energy per site of $-1/2$, does not exhibit a crossover to $\langle W_p \rangle = 1$. Instead the mean flux remains approximately zero as the temperature is lowered. This shows that frustration can lead to a degenerate manifold, where flux variables can fluctuate strongly even as $T\to 0$.\\

\begin{table}
\begin{tabular*}{\columnwidth}{l@{\hskip 0.5cm}c@{\hskip 0.75cm}c@{\hskip 0.7cm}c@{\hskip 0.7cm}c}
	\hline
	\hline
	\qquad Model &  $n$ &  $S_{0}/N_{tot}$ &  $E_{GS}/N_{tot}$ &  $\xi_{(T\to0)}$ \\
	\hline
	\quad  $S=1$ (FM) &   4  &  0.183  & -1.5  &  3.974  \\
	\quad  $S=1$ (FM) &   6  &  0.171  & -1.5  &  6.166  \\
	\quad  $S=1$ (FM) &   8  &  0.167  & -1.5  &  8.452  \\
	\quad  $S=2$ (FM) &   4  &  0.707  & -1.5  &  3.718  \\ 
	\hline
	\quad  $S=1$ (AF) &   4  &  0.837  & -0.5  &  0.846  \\
	\quad  $S=1$ (AF) &   6  &  0.837  & -0.5  &  0.848  \\
	\quad  $S=1$ (AF) &   8  &  0.837  & -0.5  &  0.848  \\
	\quad  $S=2$ (AF) &   4  &  0.520  & -1.5  &  Diverges  \\
	\hline
	\hline
\end{tabular*}
\caption{\label{tab:table-1} Transfer matrix results for the $S=1$ and $S=2$ models, for both the ferromagnetic (FM) and antiferromagnetic (AF) cases. The residual entropy $S_0$ and ground state energy $E_{GS}$ per site are given for each system width $n$ studied. The low-temperature limit of the state-state correlation length is denoted $\xi_{(T\to0)}$.}
\end{table}

\section{V. Summary and Conclusions}

In this work we have studied the properties of a modified Kitaev model introduced by BSS~\cite{baskaran} for integer spins. There is a fundamental differences between half-integer and integer spin cases: for half-integer spins, the modified model is equivalent to many copies of the original spin-1/2 Kitaev model. For example, for $S=3/2$, the $4^N$-dimensional Hilbert space decomposes into $2^N$ copies of $2^N$-dimensional Hilbert spaces, each representing a spin-half Kitaev model. Consequently, the partition function for the $S=3/2$ modified Kitaev model (and for half-integer spin in general) is equal to the partition function for the original $S=1/2$ Kitaev model multiplied by a constant. Thermodynamic quantities such as specific heat are therefore identical, with the entropy differing by just an additive temperature independent constant. 
But for integer spins, the eigenstates of our modified Kitaev model can be chosen to be a product state from site to site with no entanglement. The large degeneracy of these models, which we have explored in this work through examinations of the residual entropy, thus allows for the possibility of constructing entangled states through a superposition of the many degenerate states.

Residual entropy plateaus at intermediate temperatures are a generic feature 
of highly frustrated systems including the Kitaev systems and their 
experimental realizations~\cite{koga2, baharami,bradley,loidl,singh-oitmaa,ysingh,yamada,nasu}. These represent a classical spin-liquid state~\cite{Balents_nature}, with a highly degenerate manifold of configurations, which is a precursor to the quantum selection that leads to the quantum spin-liquid at still lower temperatures. Thus, the modified Kitaev models allow us to study such an intermediate temperature behavior with many variations. Since the integer-$S$ modified Kitaev models do not have quantum fluctuations, they are, however, missing the physics of ground-state selection and quantum mechanical excitations. 

Our study highlights that the modified Kitaev models do not just have a difference between half-integer and integer-spin cases. While all the half-integer models are closely related,
within the integer spin models we find a rich variety of behaviors, which depends on the spin value and also
on whether the coupling is ferromagnetic or antiferromagnetic. There are striking differences in their thermodynamic properties, which we have summarized in Table I. In particular, the ferromagnetic models appear critical as $T\to 0$ in the two-dimensional thermodynamic limit, with transfer matrix results indicating a correlation length scaling with the system width. Our Monte Carlo simulations identify that this diverging correlation length corresponds to state-state correlation functions. In contrast to the ferromagnetic case, for the antiferromagnetic models, the correlations remain
short-ranged as $T\to 0$ for $S=1$, whereas they diverge exponentially fast for $S=2$. This suggests that there may be an order-disorder transition at $T=0$ for antiferromagnetic models, if we regard spin as a continuous variable, between $S=1$ and $S=2$.

For all but the $S=1$ antiferromagnetic case, every bond contributes the minimum allowed energy in the ground state. With the absence of frustration in this sense, we can rigorously show that
the average flux through a hexagonal plaquette goes to $\langle W_p \rangle = 1$ as $T \rightarrow 0$. This is
confirmed by the Monte Carlo simulations. Similar result is known to hold for the spin-one Kitaev model with ferromagnetic interactions. The spin-1 antiferromagnet, however, is more strongly frustrated than others. In this case, all bonds do not contribute minimum possible energy of $-1$ to the ground state and, indeed, we find in the simulations, that the average flux through a plaquette remains close to zero even as $T \to 0$. This latter result remains to be explained analytically. 

The residual entropy per site, in the models, differs for each of the integer-spin models studied. We have provided analytical insights into the residual entropy values for the two ferromagnetic cases. For the antiferromagnetic models, one can provide bounds to the residual entropies, but their exact values have not been explained. For the spin-$S$ Kitaev models, we have given analytical arguments why the residual entropy is $S_{max}/2$, in agreement with previous numerical results~\cite{oitmaa}. When additional terms are added to higher spin Kitaev Hamiltonian~\cite{khait}, there is numerical evidence that there may still be a residual entropy, but it may not have a simple value as in the pure Kitaev case. 

Finally, we note that although the modified Kitaev models are not expected to describe real materials because of the special nature of the spin-spin interactions, they are motivated by a large body of current work exploring models with bond-dependent interactions. Also, their thermodynamic behavior shares features with spin-$S$ Kitaev models at intermediate temperatures. 
Indeed, a double peaked structure in the heat capacity and a plateau-like feature in the entropy with varying values are being discussed as experimental signatures of Kitaev physics~\cite{baharami,loidl,ysingh}. We hope that our study provides further motivation to examine higher spin Kitaev materials with both ferromagnetic and antiferromagnetic interactions \cite{stavropoulos,xu}, which this model illustrates can exhibit quite distinct thermodynamic behaviors.


\section{Acknowledgments} 
This work is supported in part by U.S.~National Science Foundation DMR Grant No.~1855111. J.O. acknowledges computing resources provided through the 
Australian Computational Infrastructure (NCI) program. D.S. thanks DST, India, 
for Project No. SR/S2/JCB-44/2010 for financial support. 

\section{Appendix A: Simultaneous diagonalization of the operators 
$\tau^\alpha$}

For integer spins, the operators $\tau^\alpha = e^{i\pi S^\alpha}$ $(\alpha = x,y,z)$ commute with each other and thus can be simultaneously diagonalized. In this section, we show that for $S=1$ the diagonalized tau operators yield three possible states per site, which have $(\tau_x, \tau_y, \tau_z) = (1, -1, -1)$, $(-1, 1, -1)$, and $(-1, -1, 1)$. We will also show that for $S=2$ there are five possible states per site, which have $(\tau^x, \tau^y, \tau^z) = (-1, -1, 1)$, $(-1, 1, -1)$, $(1, 1, 1)$, $(1, -1, -1)$, and $(1, 1, 1)$. Working in the $S_z$ basis, one may exponentiate the spin matrices for $S=1$ to obtain
\begin{equation}
e^{i\pi S_x} = \exp \left[i \pi \frac{1}{\sqrt{2}}\begin{pmatrix}
0 & 1 & 0 \\
1 & 0 & 1 \\
0 & 1 & 0 \\
\end{pmatrix}\right] = \begin{pmatrix}
0 & 0 & -1 \\
0 & -1 & 0 \\
-1 & 0 & 0 \\
\end{pmatrix} \equiv A,
\end{equation}
\begin{equation}
e^{i\pi S_y} = \exp \left[i \pi \frac{1}{\sqrt{2}i}\begin{pmatrix}
0 & 1 & 0 \\
-1 & 0 & 1 \\
0 & -1 & 0 \\
\end{pmatrix}\right] = \begin{pmatrix}
0 & 0 & 1 \\
0 & -1 & 0 \\
1 & 0 & 0 \\
\end{pmatrix} \equiv B,
\end{equation}
\begin{equation}
e^{i\pi S_z} = \exp \left[i \pi \begin{pmatrix}
1 & 0 & 0 \\
0 & 0 & 0 \\
0 & 0 & -1 \\
\end{pmatrix}\right] = \begin{pmatrix}
-1 & 0 & 0 \\
0 & 1 & 0 \\
0 & 0 & -1 \\
\end{pmatrix} \equiv C.
\end{equation}
Let
\begin{equation}V = \begin{pmatrix}
1 & 0 & -1 \\
0 & 1 & 0 \\
1 & 0 & 1 \\
\end{pmatrix}, \quad V^{-1} = \begin{pmatrix}
\frac{1}{2} & 0 & \frac{1}{2} \\
0 & 1 & 0 \\
-\frac{1}{2} & 0 & \frac{1}{2} \\
\end{pmatrix}.
\end{equation}
Then $V^{-1} A V$, $V^{-1} B V$, and $V^{-1} C V$ are all diagonal, i.e., 
\{A,B,C\} are simultaneously diagonalizable,
\begin{equation}
V^{-1} A V = \begin{pmatrix}
-1 & 0 & 0 \\
0 & -1 & 0 \\
0 & 0 & 1 \\
\end{pmatrix},
\end{equation}
\begin{equation}
V^{-1} B V = \begin{pmatrix}
1 & 0 & 0 \\
0 & -1 & 0 \\
0 & 0 & -1 \\
\end{pmatrix},
\end{equation}
\begin{equation}
V^{-1} C V = \begin{pmatrix}
-1 & 0 & 0 \\
0 & 1 & 0 \\
0 & 0 & -1 \\
\end{pmatrix}.
\end{equation}
The diagonal elements of $V^{-1} A V$, $V^{-1} B V$, and $V^{-1} C V$ yield the three states described previously (labeled $\sigma=\{1,2,3\}$ below)
\begin{equation}
\langle 1 | \tau_x | 1 \rangle = -1,~ \quad \langle 1 | \tau_y | 1 \rangle = 1,~ \quad \langle 1 | \tau_z | 1 \rangle = -1,
\end{equation}
\begin{equation}
\langle 2 | \tau_x | 2 \rangle = -1,~ \quad \langle 2 | \tau_y | 2 \rangle = -1,~\quad \langle 2 | \tau_z | 2 \rangle = 1,
\end{equation}
\begin{equation}
\langle 3 | \tau_x | 3 \rangle = 1,~ \quad \langle 3 | \tau_y | 3 \rangle = -1,~ \quad \langle 3 | \tau_z | 3 \rangle = -1.
\end{equation}
Hence for $S=1$ there are three possible states per site, which have $(\tau^x, \tau^y, \tau^z) = (1, -1, -1)$, $(-1, 1, -1)$, and $(-1, -1, 1)$. Again working in the $S_z$ basis, one can similarly exponentiate the spin matrices for $S=2$ to obtain $e^{i\pi S_x}\equiv A$, $e^{i\pi S_y}\equiv B$, and $e^{i\pi S_z}\equiv C$, which are now $5 \times 5$ matrices. 
Then one can construct the following matrix $V$ such that $V^{-1} A V$, $V^{-1} B V$, and $V^{-1} C V$ are all diagonal,
\begin{equation} V = \begin{pmatrix}
-1 & 0 & 1 & 0 & 0 \\
0 & -1 & 0 & 1 & 0 \\
0 & 0 & 0 & 0 & 1\\
0 & 1 & 0 & 1 & 0\\
1 & 0 & 1 & 0 & 0 \\
\end{pmatrix}, \quad V^{-1} = \begin{pmatrix}
-1 & 0 & 1 & 0 & 0 \\
0 & -1 & 0 & 1 & 0 \\
0 & 0 & 0 & 0 & 1\\
0 & 1 & 0 & 1 & 0\\
1 & 0 & 1 & 0 & 0 \\
\end{pmatrix}.
\end{equation}
Hence \{A,B,C\} are simultaneously diagonalizable,
\begin{equation}
V^{-1} A V = \begin{pmatrix}
-1 & 0 & 0 & 0 & 0 \\
0 & -1 & 0 & 0 & 0 \\
0 & 0 & 1 & 0 & 0\\
0 & 0 & 0 & 1 & 0\\
0 & 0 & 0 & 0 & 1 \\
\end{pmatrix},
\end{equation}
\begin{equation}
V^{-1} B V = \begin{pmatrix}
-1 & 0 & 0 & 0 & 0 \\
0 & 1 & 0 & 0 & 0 \\
0 & 0 & 1 & 0 & 0\\
0 & 0 & 0 & -1 & 0\\
0 & 0 & 0 & 0 & 1 \\
\end{pmatrix},
\end{equation}
\begin{equation}
V^{-1} C V = \begin{pmatrix}
1 & 0 & 0 & 0 & 0 \\
0 & -1 & 0 & 0 & 0 \\
0 & 0 & 1 & 0 & 0\\
0 & 0 & 0 & -1 & 0\\
0 & 0 & 0 & 0 & 1 \\
\end{pmatrix}.
\end{equation}
As before, the diagonal elements of $V^{-1} A V$, $V^{-1} B V$, and $V^{-1} C V$ give the following states:
\begin{equation}
\langle 1 | \tau_x | 1 \rangle = -1,~ \quad \langle 1 | \tau_y | 1 \rangle = -1,~ \quad \langle 1 | \tau_z | 1 \rangle = 1 
\end{equation}
\begin{equation}
\langle 2 | \tau_x | 2 \rangle = -1,~ \quad \langle 2 | \tau_y | 2 \rangle = 1,~ \quad \langle 2 | \tau_z | 2 \rangle = -1, 
\end{equation}
\begin{equation}
\langle 3 | \tau_x | 3 \rangle = 1,~ \quad \langle 3 | \tau_y | 3 \rangle = 1,~ \quad \langle 3 | \tau_z | 3 \rangle = 1, 
\end{equation}
\begin{equation}
\langle 4 | \tau_x | 4 \rangle = 1,~ \quad \langle 4 | \tau_y | 4 \rangle = -1,~ \quad \langle 4 | \tau_z | 4 \rangle = -1, 
\end{equation}
\begin{equation}
\langle 5 | \tau_x | 5 \rangle = 1,~ \quad \langle 5 | \tau_y | 5 \rangle = 1,~ \quad \langle 5 | \tau_z | 5 \rangle = 1.
\end{equation}
We therefore see that for $S=2$ there are five possible states per site, which have $(\tau^x, \tau^y, \tau^z) = (-1, -1, 1)$, $(-1, 1, -1)$, $(1, 1, 1)$, $(1, -1, -1)$, and $(1, 1, 1)$.

\appendix{}
\section{Appendix B: Argument for zero modes from spin wave theory}

In this section, we will study the large-$S$ limit of the Kitaev model defined
on a honeycomb lattice~\cite{baskaran}. We will consider a classical spin 
configuration and compute the spin-wave excitations around that configuration. 

We consider a closed loop formed by a string of bonds.
Each bond consists of nearest-neighbor spins at sites
$m$ and $n$ which interact with each other through a term of the form
$S_m^a S_n^a$, where $a$ can be $x$, $y$, or $z$; the value of $a$ must
necessarily be different for two successive bonds along the loop. Any closed
loop must have an even number of sites which alternately lie on the $A$ and 
$B$ sublattices of the system.

Given that the nearest-neighbor couplings $S_m^a S_n^a$ involve different
values of $a$ for successive bonds, we can perform a unitary transformation
to make $a=x,y,x,y,\cdots,x,y$ as we go around the closed loop. Given a 
loop formed by $2N$ sites, there are $N$ unit cells which are labeled as
$j$, and each unit cell has two sites labeled as $(j,1)$ and $(j,2)$.
If the sites have spin-$S$, the Hamiltonian for the loop is given by
\begin{equation} H ~=~ \frac{J}{S} ~\sum_{j=1}^N ~(S_{j,1}^x S_{j,2}^x ~+~ 
S_{j,2}^y S_{j+1,1}^y), \label{ham1} \end{equation}
and we have a periodic boundary condition for $i$. We assume that $J > 0$. (If 
$J < 0$, we can change its sign by performing a unitary rotation which flips 
the signs of $S_{j,1}^x$, $S_{j,1}^z$, $S_{j,2}^y$, and $S_{j,2}^z$ for all 
values of $j$). A factor of $1/S$ has been introduced in Eq.~\eqref{ham1} so 
that the ground-state energy is proportional to $S$ in the limit $S \to 
\infty$.

We will now use the Holstein-Primakoff (HP) 
transformation~\cite{holstein,anderson,kubo} to compute the spin-wave spectrum 
for a closed loop. We assume that the classical spin configuration is such that
the spin at a site $n$ lying on the loop points along either the direction 
$+ \hat a$ or $- \hat a$, where $\hat a$ is different from the directions of 
the interactions of that site with its two nearest neighbors along the 
loop. For instance, if site $n$ interacts with site $n-1$ with an interaction
$S_n^x S_{n-1}^x$ and with site $n+1$ with interaction $S_n^y S_{n+1}^y$,
then classically the spin at site $n$ is given by ${\vec S}_n = \pm S 
{\hat z}$. Let us now consider these two cases separately.

\noindent (i) If the spin at site $n$ points along the $\hat z$ direction, then
we have
$S_n^z = S - (p_n^2 + q_n^2 -1)/2$, and $(S_n^x,S_n^y)$ can be chosen in four 
different ways, namely, $\sqrt{S} (q_n,p_n), \sqrt{S} (-q_n,-p_n), \sqrt{S} 
(p_n,-q_n)$, and $\sqrt{S} (-p_n,q_n)$, up to the lowest order in the HP 
transformation. Here $q_n$ and $p_n$ are canonically conjugate variables
which satisfy the commutation relation $[q_n,p_n] = i$.

\noindent (ii) If the spin points along the $-\hat z$ direction, then we have $S_n^z
= - S + (p_n^2 + q_n^2 -1)/2$, and $(S^x,S^y)$ can be chosen in four ways, 
namely, $\sqrt{S} (p_n,q_n), \sqrt{S} (-p_n,-q_n), \sqrt{S} (q_n,-p_n)$ and 
$\sqrt{S} (-q_n,p_n)$.

We now compute the spin-wave spectrum for a closed loop with $2N$ sites.
(The minimum value of $2N$ is 6 corresponding to a hexagon). 
As we go around the loop, we choose 
the spin variables along the loops to be $q$ and $p$ alternately, so that the 
couplings between nearest neighbors involve either $(q_m,q_n)$ or $(p_m,p_n)$ 
but not $(q_m,p_n)$. Because of the two cases (i) and (ii) discussed
above, the loop may have either periodic boundary condition (PBC) or 
antiperiodic boundary condition (ABC) depending on the set of classical
spin directions ${\vec S}_n$ as we go around the loop. Ignoring some constants, 
we find the spin-wave Hamiltonian for the loop to be
\begin{eqnarray} H_{sw} &=& \frac{J}{2} ~\sum_{n=1}^{2N} ~(p_n^2 ~+~ q_n^2)~ 
\nonumber \\
& & +~ J~ \sum_{n=1}^{N-1} ~(p_{2n-1} p_{2n} ~+~ q_{2n} q_{2n+1}) \nonumber \\
& & +~ J~ (p_{2N-1} p_{2N} ~\pm~ q_{2N} q_1), \label{ham2} \end{eqnarray}
with either PBC or ABC for the last bond connecting sites $2N$ and $1$; the 
sign of term $q_{2N} q_1$ term is $+$ and $-$ in the two cases, respectively. 
Solving for the spectrum, we find that the normal modes can be 
characterized by a momentum $k$, where $k = 0, ~2\pi /N, ~\cdots , ~(2\pi N - 
2\pi)/N$ in the case of PBC, and $k = \pi /N, ~3\pi /N, ~\cdots , ~(2\pi N - 
\pi)/N$ in the case of ABC; in each case, $k$ can take $N$ different values. 
We now find the normal mode frequencies by solving
the classical Hamiltonian equations of motion. (For a quadratic Hamiltonian,
the frequencies turn out to be the same regardless of whether we study the 
problem classically or quantum mechanically). For each momentum $k$, we find 
that there are two frequencies given by 0 and $\omega_k = 2 J |\cos (k/2)|$.
The existence of a zero-energy mode for each $k$ implies a large 
ground-state degeneracy for the following reason. Since the total number
of states of the loop is $(2S+1)^{2N}$, we can take each of the $2N$ modes
to describe $2S+1$ possible states. The $N$ zero modes therefore imply that
there are $(2S+1)^N$ states all of which have zero energy and are therefore
degenerate. This seems to agree well with the numerical results reported
in Ref.~\onlinecite{oitmaa}, even for cases where $S$ is not very large.
Namely, for $S=1, 3/2$ and $2$, it is found that there is a low-energy manifold
with an entropy per site given by $(1/2) \ln (2S+1)$.

\end{document}